\documentclass[12pt]{article}

\def\tr{{\rm tr}}

\begin{document}

\title{Intermediate inflation and the slow-roll approximation}

\author{Alan D. Rendall\\
Max-Planck-Institut f\"ur Gravitationsphysik\\Albert-Einstein-Institut
\\ Am M\"uhlenberg 1\\
14476 Golm, Germany}

\date{}

\maketitle

\begin{abstract}
It is shown that spatially homogeneous solutions of the Einstein 
equations coupled to a nonlinear scalar field and other matter exhibit 
accelerated expansion at late times for a wide variety of potentials $V$. 
These potentials are strictly positive but tend to zero at infinity. 
They satisfy restrictions on $V'/V$ and $V''/V'$ related to
the slow-roll approximation. These results generalize Wald's 
theorem for spacetimes with positive cosmological constant to
those with accelerated expansion driven by potentials belonging to a large
class.
\end{abstract}

\section{Introduction}

In recent years astronomical observations have shown the necessity of
incorporating accelerated expansion into the description of the evolution
of our universe. Unfortunately there is no clear understanding of the 
mechanism leading to the acceleration. The simplest possibilities are a
cosmological constant or a nonlinear minimally coupled scalar field with
a potential. There is now a large and growing literature on nonlinear
scalar fields with different potentials and many more exotic models.
The observational data available at present only gives weak criteria
for deciding between different models. For instance it is not even clear 
that it can rule out the case of a cosmological constant (cf. \cite{alam}).
It is to be hoped that this situation will improve due the accumulation of 
more data in the next few years. In the meantime it makes sense to advance
our theoretical understanding of the models available. 

The present paper is concerned with the dynamics of spatially homogeneous
solutions of the Einstein equations in the presence of a minimally coupled 
nonlinear scalar field and other matter. The latter represents ordinary
baryonic matter and non-baryonic dark matter. In the literature on 
models with accelerated cosmological expansion one or more of the 
following effects are often ignored: shear, ordinary matter, spatial 
curvature. The usual intuitive picture says that they will be negligible
under suitable circumstances in a phase of accelerated expansion. A
rigorous proof that this is true was given by Wald \cite{wald} in the
case the acceleration is caused by a cosmological constant. The first
main result of the present paper is that Wald's theorem can be generalized 
to the case where the cosmological constant is replaced by a scalar field 
belonging to a large class. This class apparently includes all of
the potentials considered in the literature for which accelerated expansion 
is expected to continue indefinitely.

A concept which is frequently used when studying the dynamics of 
models with accelerated expansion is that of the slow-roll approximation.
It says that under certain circumstances the equation of motion of
the scalar field can be replaced by a simpler equation while retaining
the essential qualitative features of the evolution. The precise
meaning of the slow-roll approximation has been investigated in
\cite{liddle}. It was shown that if a solution approaches the 
\lq slow-roll attractor\rq\ many conclusions follow. The second main
result of the present paper is a proof that for a wide class of potentials
all solutions approach the slow-roll attractor. 

Since the potential of the scalar field leading to accelerated expansion
is not known it is useful to try to analyse the dynamics of 
spacetimes with a nonlinear scalar field for large classes of potentials
assumed only to satisfy some qualitative restrictions. This programme
was carried out in \cite{rendall04a} for potentials with a strictly
positive lower bound and solutions of Bianchi types I-VIII. The spacetimes
analysed may contain matter satisfying the dominant and strong energy 
conditions in addition to the scalar field and the results are a direct
generalization of Wald's theorem \cite{wald} on spacetimes with positive
cosmological constant. As pointed out in \cite{rendall04a} the next natural 
case to look at in order of increasing difficulty is that where the 
potential is everywhere positive but may tend to zero as the scalar field 
$\phi$ tends to $\pm\infty$. Furthermore it should be assumed that the 
fall-off at infinity is not too fast, since otherwise accelerated expansion 
may not occur.

To see the need for a restriction on the fall-off of the potential it
is useful to look at explicit solutions given by Halliwell 
\cite{halliwell}. These solutions are homogeneous and isotropic and
spatially flat. Their only matter content is a scalar field with an
exponential potential. In notation which is convenient for the 
following the potential is given by $V(\phi)=V_0 e^{-\sqrt{8\pi}\lambda\phi}$,
where $\lambda$ is a positive constant. If these solutions are written in
Gauss coordinates the spatial metric is of the form 
$g_{ij}(t)=a^2(t)\delta_{ij}$ where the scale factor $a(t)$ is proportional 
to $t^{2/\lambda^2}$. The expansion is accelerated iff the power occurring is
greater than one, i.e. iff $\lambda<\sqrt2$. Thus if the aim is to consider 
models with accelerated expansion an assumption should be made on the potential
which rules out too rapid exponential decay. This can be done by means of a
restriction on $V'/V$. 
 
Acceleration due to an exponential potential is often called power-law 
inflation since the scale factor has a power-law behaviour in the simplest
cases. A class of models where the scale factor behaves in a way 
intermediate between a power law and the exponential expansion resulting
from a positive cosmological constant is associated with name intermediate
inflation \cite{barrow90a}. The specific case considered in \cite{barrow90a}
is where the scale factor is proportional to $\exp(t^{f})$ for $0<f<1$.
This can be obtained from a potential which looks asymptotically like a
negative power but is not exactly a power. Note that if a cosmological 
constant is thought of as a constant potential the potential in 
intermediate inflation is intermediate between the cosmological constant 
and the potential for power-law inflation as far as its asymptotic
behaviour for large $\phi$ is concerned. The aim of this paper is to 
analyse the case of general potentials which are positive, tend to zero as 
$\phi\to\infty$ and are such that $-V'/V$ satisfies an upper bound which
rules out exponential potentials with $\lambda\ge\sqrt{2}$. 

Under an additional assumption on $V''/V'$ further information is obtained
on the asymptotic behaviour of the solutions. These properties are related
to the slow-roll approximation \cite{liddle}. It is shown that the results 
obtained are general enough to apply to many potentials considered in the 
literature.

After some necessary background material has been presented in section 2,
the slow-roll approximation is discussed in section 3 for spatially flat
isotropic models with a non-linear scalar field as the only matter
content. Section 4 deals with the generalization to all models of 
Bianchi types I-VIII and to the case where arbitrary matter fields 
satisfying the dominant and strong energy conditions are present in
addition to the scalar field. In section 5 it is shown that the results
which have been obtained are applicable to many potentials considered in the 
literature. The late-time dynamics of perfect fluids and collisionless
matter is treated in section 6. Section 7 sums up what has been achieved
and indicates some avenues for future progress.

\section{Background} 

This section recalls the equations which are needed in the following and
some basic facts which were proved in \cite{rendall04a}. Consider a spacetime 
with vanishing cosmological constant which contains a nonlinear scalar field 
and some other matter. Suppose that the other matter satisfies the dominant 
and strong energy conditions. The energy momentum tensor is 
\begin{equation}
T_{\alpha\beta}=T^M_{\alpha\beta}+\nabla_\alpha\phi\nabla_\beta\phi
-\left[\frac12\nabla^\gamma\phi\nabla_\gamma\phi+V(\phi)\right]
g_{\alpha\beta}.
\end{equation}
where $T^M_{\alpha\beta}$ is the energy-momentum tensor of the matter other
than the scalar field. Assume that the potential $V$ is $C^2$ and non-negative.
Restricting to a Bianchi spacetime leads to the following basic equations:
\begin{eqnarray}
\frac{dH}{dt}&=&-4\pi \dot\phi^2-\frac12\sigma_{ab}\sigma^{ab}+\frac16 R
-4\pi(\rho^M+\frac13\tr S^M)    \\
\ddot\phi&=&-3H\dot\phi-V'(\phi)\label{evphi}
\\
H^2&=&\frac{4\pi}3[\dot\phi^2+2V(\phi)]+\frac16(\sigma_{ab}\sigma^{ab}-R)
+\frac{8\pi}3\rho^M\label{ham}
\end{eqnarray}
Here $H$ is the Hubble parameter, $\sigma_{ab}$ is the tracefree part of
the second fundamental form, $R$ is the scalar curvature of the spatial
metric and $\rho^M$ and $\tr S^M$ are projections of the energy-momentum
tensor $T_{\alpha\beta}^M$. For Bianchi types I to VIII the inequality
$R\le 0$ holds. Assuming that $H>0$ at some time (so that the model is 
expanding at that time) it follows that $H>0$ at all times. For more 
details see \cite{rendall04a}.

If $H$ is positive at some time then it is everywhere positive and
non-increasing and so tends to some limit $H_1\ge 0$. 
In Theorem 1 of section 3 of \cite{rendall04a} some general results on 
late-time dynamics for the above equations were proved unter the following
three assumptions:
\begin{enumerate}
\item $V(\phi)\ge V_0$ for a constant $V_0>0$
\item $V'$ is bounded on any interval on which $V$ is bounded
\item $V'$ tends to a limit, finite or infinite, as $\phi$ tends
to $\infty$ or $-\infty$.
\end{enumerate} 
Suppose for a moment that the first of these assumptions is dropped but 
the other two are retained. The first part of the proof of Theorem 1 of 
\cite{rendall04a} does not use assumption 1. and so it can still be applied.
The conclusion is that if a spacetime of the type described above exists
globally in the future then $\dot\phi\to 0$ and $V(\phi)$ tends to a limit
$V_1\ge 0$ as $t\to\infty$. If $H_1>0$ then the rest of the 
proof of the theorem also goes through and the same conclusions are
obtained as in that theorem. In particular the expansion is accelerated at
late times, the model isotropizes and the contribution of matter and spatial
curvature to the field equations becomes negligible. Moreover 
$3H^2/8\pi V\to 1$ and $V'(\phi)\to 0$ as $t\to\infty$.

It remains to consider the case $H_1=0$. In that case it follows that 
$\dot\phi\to 0$ and $V(\phi)\to 0$ as $t\to\infty$ without having to make 
the additional assumptions 2. and 3. When $H_1=0$ it follows that  
$\sigma^{ab}\sigma_{ab}$, $R$ and $\rho^M$ converge to zero 
as $t\to\infty$. The quantity 
\begin{equation}\label{zdef}
Z=9H^2-24\pi[\dot\phi^2/2+V(\phi)]
=\frac32(\sigma_{ab}\sigma^{ab}-R)+24\pi\rho^M
\end{equation}
satifies the inequality $dZ/dt\le -2HZ$. In \cite{rendall04a} this was 
combined with the fact that $H$ has a positive lower bound to show that
$Z$ decayed exponentially. In the present more general case this is no 
longer true but the differential inequality can be used in a different 
way. In fact $H^2\ge Z/9$ and so $dZ/dt\le -\frac23 Z^{3/2}$. It follows that 
$Z=O(t^{-2})$ and this gives decay rates for $\sigma^{ab}\sigma_{ab}$,
$R$ and $\rho^M$. Note that the conclusions up to this point also hold in
the absence of a scalar field. Assume now that $V$ is positive. (This 
assumption will be maintained for the rest of the paper.) Then 
$V(\phi(t))\to 0$ and so $\phi\to\infty$ or $\phi\to -\infty$ as $t\to\infty$. 
These cases can be transformed into each other by reversing the sign of $\phi$ 
and so it can be assumed without loss of generality that $\phi\to\infty$. In 
this case $V(\phi)\to 0$ as $\phi\to\infty$. 

\section{The slow-roll approximation} 

In this section we specialize to the isotropic and spatially flat case with
a nonlinear scalar field as the only matter. This is the situation most
frequently considered in the literature. In this case $H=\dot a/a$ where 
$a$ is the scale factor. It will be shown that certain 
assumptions on the potential $V$ imply accelerated expansion at late times. 
The assumptions are
\begin{enumerate}
\item $V(\phi)>0$ with $V(\phi)\to 0$ as $\phi\to\infty$
\item $V'(\phi)<0$ 
\item $V'(\phi)/V(\phi)\to 0$ as $\phi\to\infty$.
\end{enumerate}

Consider a situation where, as in the last section, $\phi\to\infty$ as 
$t\to\infty$. Note that $\dot\phi>0$ for $t$ sufficiently large. For 
$\dot\phi$ must be positive at some time. After that it must remain 
positive since $\dot\phi=0$ and assumption 2. above imply that $\ddot\phi>0$.
When it is known that $\dot\phi>0$ the assumptions of the last section can 
be weakened while maintaining the same conclusions. Assumption 2. only needs
to be required for intervals whose left endpoint is finite. In assumption
3. only the existence of the limit for $t\to\infty$ need be required.
These weakened versions of assumptions 2. and 3. of the last section follow 
from the assumptions 1.-3. of this section.
The following argument was inspired by a paper of Muslimov \cite{muslimov}
on the isotropic case but we will not use the complicated transformation of 
variables of that paper directly. Since $\phi$ is monotone we can regard
functions of $t$ as functions of $\phi$. The key relation is the following:
\begin{equation}\label{phider}
\frac{d}{d\phi}\left(\frac{3H^2}{8\pi V}\right)=\frac{3H^2}{8\pi V}
\left(-\frac{8\pi\dot\phi}{H}-\frac{V'}{V}\right)
\end{equation}

\noindent
{\bf Theorem 1} Consider a spatially flat homogeneous and isotropic 
solution of the Einstein equations coupled to a nonlinear scalar field with 
potential $V$ of class $C^2$ satisfying conditions 1. - 3. above.
Suppose that the solution is initially expanding ($H>0$) and that 
$\dot\phi>0$ at some time. If the solution exists globally to the 
future then $3H^2/8\pi V(\phi)\to 1$ as $t\to\infty$ and $\ddot a>0$ for
$t$ sufficiently large.

\noindent
{\bf Proof} In order to control the first term on the right hand side of 
(\ref{phider}) note that 
\begin{equation}\label{ratio}
\frac{\dot\phi}{H}=\sqrt{\frac{3}{4\pi}-\frac{2V(\phi)}{H^2}}
\end{equation}
and that the right hand side is an increasing function of $3H^2/8\pi V$.
Moreover it tends to zero as $3H^2/8\pi V\to 1$. 
Let $C_1>8\pi/3$ be a constant and let $C_2=8\pi\sqrt{3/4\pi-2/C_1}$. 
Since $V'/V\to 0$ as $\phi\to\infty$ there exists some $\phi_1$ such that 
$-V'(\phi)/V(\phi)\le C_2/2$ for all $\phi\ge \phi_1$. At any point where
$\phi\ge\phi_1$ and $H^2/V\ge C_1$ the expression in brackets on the right 
hand side of (\ref{phider}) is bounded above by $-C_2/2$. Hence 
$H^2/V$, if it is ever greater than $C_1$, reaches $C_1$ at some greater 
value $\phi_2$ of $\phi$ and remains below that value for all $\phi\ge\phi_2$.
Because of the freedom to choose $C_1$, and the fact that $3H^2/8\pi V\ge 1$, 
it follows that $3H^2/8\pi V\to 1$ as $\phi\to\infty$ and hence as 
$t\to\infty$. Thus the first conclusion of the theorem has been proved.

The inequality $\dot H+H^2\ge 0$ is 
equivalent to $\ddot a>0$ and is thus the criterion for accelerated expansion. 
Substituting for $H$ and $\dot H$ using the field equations shows that it is 
equivalent to the condition that $3H^2/8\pi V(\phi)<3/2$, which is fulfilled 
at late times, giving the remaining conclusion of the theorem.

\vskip 10pt
\noindent
{\bf Remark} The inequality $\dot H+H^2\ge 0$ implies that 
$H(t)\ge (t+C)^{-1}$ for some $C>0$ and $t$ large and that 
$\int_{t_0}^\infty H(t) dt=\infty$ for any $t_0$.

An analogous result can be obtained in the case that assumption 3. above is
replaced by the weaker condition that
$\limsup (-V'/V)\le\alpha$ for a suitable constant $\alpha$. 

\noindent
{\bf Theorem 2} Consider a spatially flat homogeneous and isotropic 
solution of the Einstein equations coupled to a nonlinear scalar field with 
potential $V$ of class $C^2$ satisfying conditions 1. - 2. above
with $\alpha=\limsup (-V'/V)<4\sqrt\pi$. Suppose that the 
solution is initially expanding ($H>0$) and that $\dot\phi>0$ at some time. 
If the solution exists globally to the future then  $\ddot a>0$ for $t$ 
sufficiently large.

\noindent
{\bf Proof} Suppose that $C_1>\beta$ where 
\begin{equation}
\frac{2}{\beta}=\frac{3}{4\pi}-\frac{\alpha^2}{64\pi^2}
\end{equation}
and define $C_2$ in terms of $C_1$ as above. Then $C_2>\alpha$.
There exists some $\phi_1$ such that $-V'/V\le (C_2+\alpha)/2$ for 
$\phi>\phi_1$. If $\phi\ge\phi_1$ and $H^2/V\ge C_1$ then 
$-8\pi\dot\phi/H-V'/V\le (\alpha-C_2)/2<0$. It follows that, as in
the previous case, $(H(\phi))^2/V(\phi)$, if it is ever greater than $C_1$, 
reaches $C_1$ at some greater value $\phi_2$ of $\phi$ and remains below that 
value for all $\phi\ge\phi_2$. Hence in the limit $\phi\to\infty$ we have 
$\limsup 3H^2/8\pi V\le 3\beta/8\pi$. 
Thus the condition for accelerated expansion is $3\beta/8\pi<3/2$, i.e. 
$\alpha <4\sqrt{\pi}$. In particular the inequality for accelerated 
expansion in the case of an exponential potential with 
$\lambda=\alpha/\sqrt{8\pi}$ is recovered.

\vskip 10pt
\noindent
{\bf Remark} For an exponential potential it follows from \cite{lee04b}
that $3H^2/8\pi V\to 3\beta/8\pi$.

\vskip 10pt\noindent
In the case that $V'/V\to 0$ as $t\to\infty$ it is possible to obtain
further information on the asymptotic behaviour of the solutions which
relates to the so-called slow-roll approximation \cite{liddle}. 
This says that under the assumption on the potential that $V''/V'$ is
bounded the term $\ddot\phi$ in the equation of motion of the scalar field 
becomes negligible at late times. Equivalently $3H\dot\phi/V'\to -1$. In the 
terminology of \cite{liddle} this shows that for a potential satisfying this 
assumption all solutions approach the slow-roll attractor. 

\noindent
{\bf Theorem 3} Let the assumptions of Theorem 1 hold and assume in addition
that $V''/V'$ is bounded as $\phi\to\infty$. Then $3H\dot\phi/V'\to -1$
and $\ddot\phi/3H\dot\phi\to 0$ as $t\to\infty$.

\noindent
{\bf Proof} The desired 
limiting behaviour can be proved using the relation 
\begin{equation}\label{slowroll}
\frac{d}{dt}\left(\frac{H\dot\phi}{V'}\right)=-H\left[\left(1+\frac{3H\dot\phi}
{V'}\right)+\frac{H\dot\phi}{V'}\left(\frac{4\pi\dot\phi^2}{H^2}
+\frac{\dot\phi}{H}\frac{V''}{V'}\right)\right]
\end{equation}
Let $C_1>0$ be a constant. There exists a time $t_1$ such that for $t\ge t_1$
the inequality $(\dot\phi/H)(V''/V')\ge -C_1/(1+C_1)$ holds. If 
$3H\dot\phi/V'\le -1-C_1$ at some time $t\ge t_1$ then it follows from
(\ref{slowroll}) that 
\begin{equation}
\frac{d}{dt}\left(\frac{H\dot\phi}{V'}\right)\ge -\frac{2C_1H}{(1+C_1)}
\left(\frac{H\dot\phi}{V'}\right).
\end{equation} 
Since $H$ is
not integrable it follows that $1+3H\dot\phi/V'$ must reach $-C_1$ after 
finite time and cannot become less than this value again. Using the freedom 
to choose $C_1$ shows that $\liminf (1+H\dot\phi/V')\ge 0$. In particular,
since $V'$ is negative, it follows that $H\dot\phi/V'$ is bounded.
Now choose $t_2$ such that the modulus 
of the second term in the square bracket in (\ref{slowroll}) is less than 
$C_1/2$ for $t\ge t_2$. Suppose that at some time $t\ge t_2$ the inequality
$1+3H\dot\phi/V'\ge C_1$ holds. Then $1+3H\dot\phi/V'$ is decreasing there.
Moreover the rate of decrease is such it must reach $C_1$ in finite time.
It follows that $1+3H\dot\phi/V'\to 0$ as $t\to\infty$. The equation of 
motion for $\phi$ implies that $\ddot\phi/3H\dot\phi\to 0$. 

\vskip 10pt
\noindent
{\bf Remark} For an exponential potential it follows from \cite{lee04b}
that $3H\dot\phi/V'\to -(1-\lambda^2/6)^{-1}$ as $t\to\infty$.

\section{The anisotropic case with matter}

Here we generalize the results of the last section to anisotropic solutions 
with matter. 

\noindent
{\bf Theorem 4} Consider a solution of the Einstein equations of Bianchi
type I-VIII coupled to a nonlinear scalar field with potential $V$
of class $C^2$ satisfying conditions 1. - 3. of section 3 and other matter 
satisfying the dominant and strong energy conditions. Suppose that the 
solution is initially expanding ($H>0$) and that $\dot\phi>0$ at some time.
Then

\noindent
a) if the solution exists globally to the future then
$3H^2/8\pi V(\phi)\to 1$, $\sigma_{ab}\sigma^{ab}/H^2$, $R/H^2$, and
$\rho^M/H^2$ tend to zero as $t\to\infty$ and $\dot H+H^2\ge 0$ for $t$ 
sufficiently large and

\noindent
b) if in addition $V''/V'$ is bounded then $3H\dot\phi/V'\to -1$ as
$t\to\infty$.

\noindent
{\bf Proof} Equation (\ref{phider}) is replaced by
\begin{eqnarray}\label{phidergen}
\frac{d}{d\phi}\left(\frac{3H^2}{8\pi V}\right)&=&\frac{3H^2}{8\pi V}
\left[-\frac{8\pi\dot\phi}{H}-\frac{V'}{V}\right. \nonumber  \\
&&\left.-\frac{2}{H\dot\phi}\left(\frac12\sigma_{ab}\sigma^{ab}-\frac16 R
+4\pi\left(\rho^M+\frac13\tr S^M\right)\right)\right]
\end{eqnarray}
while (\ref{ratio}) is replaced by
\begin{equation}\label{ratiogen}
\frac{\dot\phi}{H}=\left[{\frac{3}{4\pi}-\frac{2V(\phi)}{H^2}}
-\frac{1}{8\pi}\frac{\sigma_{ab}\sigma^{ab}-R}{H^2}
-\frac{2\rho^M}{H^2}\right]^{1/2}
\end{equation}
It will be shown that (\ref{phidergen}) implies that $\dot\phi/H\to 0$ as 
$\phi\to\infty$. Given a constant $C_1>0$ there exists $\phi_1$ such that 
$V'/V\le C_1/2$ for all $\phi\ge \phi_1$. Hence if $\dot\phi/H\ge C_1/8\pi$ 
and $\phi\ge \phi_1$ then $d/d\phi(3H^2/8\pi V)\le -(C_1/2)(3H^2/8\pi V)$. 
Thus eventually 
$\dot\phi/H$ reaches the value $C_1$ and does not return. This proves the 
desired result. Now 
\begin{eqnarray}
&&-\frac{8\pi\dot\phi}{H}-\frac{2}{H\dot\phi}\left(
\frac12\sigma_{ab}\sigma^{ab}-\frac16 R+4\pi\left(\rho^M+\frac13\tr S^M\right)
\right)
\nonumber\\
&&\le -\sqrt{48\pi}+\sqrt{48\pi}\left(\frac{8\pi V(\phi)}{3H^2}\right)+
\frac{\sigma_{ab}\sigma^{ab}}{H^2}\left(\frac16\sqrt{48\pi}-\left(
\frac{\dot\phi}{H}\right)^{-1}\right)
\\
&&-\frac{R}{H^2}\left(\frac16\sqrt{48\pi}-\frac13\left(\frac{\dot\phi}{H}
\right)^{-1}\right)+\frac{8\pi}{3H^2}\left(\sqrt{48\pi}\rho^M-
\left(
\frac{\dot\phi}{H}\right)^{-1}\left(3\rho^M+\tr S^M\right)\right)\nonumber
\end{eqnarray} 
Provided $\dot\phi/H\le 1/\sqrt{12\pi}$ then
\begin{equation}
\frac{d}{d\phi}\left(\frac{3H^2}{8\pi V}\right)\le -\left[\sqrt{48\pi}
\left(1-\frac{8\pi V(\phi)}{3H^2}\right)-\frac{V'}{V}\right]
\left(\frac{3H^2}{8\pi V}\right)
\end{equation}
Thus arguing
as in the isotropic case shows that $3H^2/8\pi V\to 1$ as $\phi\to\infty$,
i.e. as $t\to\infty$. It follows that $\sigma_{ab}\sigma^{ab}/H^2$,
$R/H^2$ and $\rho^M/H^2$ tend to zero as $t\to\infty$. In particular 
$\dot H+H^2\ge 0$ at late times and there is accelerated expansion. 

The argument to obtain the slow-roll approximation works just as in the
isotropic case. Equation (\ref{slowroll}) is changed by replacing
$4\pi\dot\phi^2$ in the last bracket by 
\begin{equation}
H^{-2}\left[ 4\pi\dot\phi^2+\frac12 \sigma_{ab}\sigma^{ab}-\frac16 R
+4\pi\left(\rho^M+\frac13\tr S^M\right)\right]
\end{equation}
The extra terms which are added have the right sign and decay as 
$t\to\infty$ and this is all that is needed.

\vskip 10pt
With some more work this result can be strengthened. Let $\epsilon>0$. 
It follows from the theorem that
\begin{equation}
\dot H+\epsilon H^2=H^2(\epsilon-4\pi\dot\phi^2/H^2+o(1))
\end{equation}
Using what is known about $\dot\phi/H$ shows that for $t$ large the right hand
side is positive. Thus a differential inequality is obtained which implies 
that $H\ge \epsilon^{-1}(C+t)^{-1}$ for a constant $C>0$. Thus $H$ falls
off slower than any positive multiple of $t^{-1}$. It follows from the
inequality $dZ/dt\le -2HZ$ that $Z$ decays faster than any power of $t$.
Hence $Z/H^2$ decays faster than any power of $t$. In particular
$\sigma_{ab}\sigma^{ab}/H^2$ decays faster than any power of $t$.

\section{Applications to potentials in the literature}

In the literature on inflation and quintessence many choices of potential
have been considered and we will not attempt to give a comprehensive
survey. Large classes of potentials where the theorems of this paper are
applicable will be identified and a number of examples in the literature
belonging to these classes will be pointed out. Then it will be shown
by example how the theorems can be used to prove asymptotic expansions
for the solutions. 

Consider the following class of potentials:
\begin{equation}\label{pot1}
V(\phi)=V_0(\log\phi)^p\phi^n\exp(-\lambda\phi^m)
\end{equation}
The case $p=0$ was studied in \cite{parsons1} while the cases $m=0$ and
$n=0$, $m=1$ were studied in \cite{parsons2}. The requirement that
$V\to 0$ as $\phi\to\infty$ leads to some restrictions. Suppose first 
that $\lambda<0$. Then it must be assumed that $m\le 0$. If $m=0$ it 
must further be assumed that $n\le 0$ and if $n=0$ it must be assumed
that $p<0$. If $\lambda>0$ it must be assumed that $m>0$ or that 
$n$ and $p$ satisfy the restrictions already listed. If $\lambda=0$
then $n$ and $p$ must also satisfy those restrictions. To compare with 
the other hypotheses of the theorems the first and second 
derivatives of $V$ must be computed.
\begin{eqnarray}
V'(\phi)&&=\left(\frac{p}{\phi\log\phi}+\frac{n}{\phi}-\lambda m
\phi^{m-1}\right)V      \\
V''(\phi)&&=\left[\frac{p(p-1)}{\phi^2(\log\phi)^2}+\frac{p(2n-1)}
{\phi^2\log\phi}+\frac{n(n-1)}{\phi^2}\right. \nonumber\\
&&-\lambda m\left(\frac{2p}{\phi\log\phi}
\left.+\frac{m+2n-1}{\phi}\right)\phi^{m-1}+\lambda^2m^2\phi^{2(m-1)}\right]V
\end{eqnarray}
Evidently $V'/V$ tends to zero as $\phi\to\infty$ iff $m<1$. When $m=1$
the quantity $-V'/V$ converges to a positive limit. It can easily
be checked that under the restrictions already made $V'$ is negative for
$\phi$ sufficiently large. In the case that $0<m<1$ and $\lambda>0$ it
follows that $V''/V'$ behaves asymptotically like $-\lambda m\phi^{m-1}$
and so Theorem 4 is applicable. In the case
where $m\le 0$, $n<0$ the quantity $V''/V'$ behaves asymptotically like
$(n-1)/\phi$ and so Theorem 4 again applies. The case where $m\le 0$,
$n=0$ and $p<0$ is similar, since there $V''/V'$ behaves asymptotically 
like $(p-1)/\phi\log\phi$. When $m=1$ the assumptions of Theorem 4 are
not satisfied but it follows from Theorem 2 that in the spatially flat 
isotropic case with scalar field alone the models exhibit accelerated 
expansion.

Suppose now that $W$ is a potential satisfying some of the assumptions of the 
theorems and let $V$ be a potential for which $V=W(1+o(1))$, $V'=W'(1+o(1))$ 
and $V''=W''(1+o(1))$ as $\phi\to\infty$. Then $V$ satisfies the corresponding
assumptions. In particular $\limsup (-V'/V)=\limsup (-W'/W)$. As an example, 
consider the potentials for intermediate inflation given in 
\cite{barrow90a}. These are of the form:
\begin{equation}
V(\phi)=V_0(\phi-\phi_0)^n+V_1(\phi-\phi_0)^{n-2}
\end{equation}
for constants $n$, $\phi_0$, $V_0$ and $V_1$ with $n$ negative and 
$V_0$ positive. Taking $W=V_0\phi^n$ allows the above observation to be 
applied in this case to see that Theorem 4 applies. Another example where
this procedure works is in the case $V=V_0(\exp M_p/\phi-1)$ \cite{zlatev}.
Take $W=V_0M_p/\phi$. The case where 
$\limsup (-V'/V)$ is positive gives a proof of late-time accelerated 
expansion for spatially flat isotropic models with scalar field alone
in a number of cases. There follow some examples taken from
a list of potentials in \cite{sahni}. If $V=V_0(\cosh\lambda\phi-1)^p$
\cite{sahniwang} with $-4\sqrt{\pi}<p\lambda<0$ then take 
$W=(V_0/2^{p})e^{p\lambda\phi}$. If $V=V_0\sinh^{-\alpha}(\lambda\phi)$
\cite{urena} with $0<\alpha\lambda<4\sqrt{\pi}$ take 
$W=(2^\alpha V_0)e^{-\alpha\lambda\phi}$. If $V=V_0(e^{\alpha\kappa\phi}
+e^{\beta\kappa\phi})$ \cite{barreiro} with $-4\sqrt{\pi}<\alpha\kappa<0$ 
and $\beta\kappa<\alpha\kappa$ then take $W=V_0 e^{\alpha\kappa\phi}$.
If $V=V_0[(\phi-B)^\alpha+A]e^{-\lambda\phi}$ \cite{albrecht} and $\lambda>0$ 
then take $W=V_0\phi^\alpha e^{-\lambda\phi}$ for $\alpha>0$ and 
$W=V_0Ae^{-\lambda\phi}$ for $\alpha<0$.

In \cite{parsons1} asymptotic expansions were written down for solutions 
corresponding to certain potentials. With the theorems of this paper in
hand these asymptotic expansions can be given an interpretation which can
be proved rigorously. For simplicity only the case $p=0$, $n=0$, $0<m<1$, 
$\lambda=1$ of (\ref{pot1}) will be discussed. Under the hypotheses of 
Theorem 4, with this choice of potential, $3H^2/8\pi V\to 1$ and 
$3H\dot\phi/V'\to-1$ as $t\to\infty$. It can be concluded that the solutions 
satisfy
\begin{equation}
\dot\phi=-\frac{V'}{\sqrt{24\pi V}}+o\left(\frac{V'}{\sqrt{V}}\right) 
\end{equation}
as $t\to\infty$. This should be compared with equation (2.9) of 
\cite{parsons1}. Substituting the specific form of the potential gives
\begin{equation}
\dot\phi=m\sqrt{\frac{V_0}{24\pi}}\phi^{m-1}\exp \left(-\frac12\phi^m\right)
+o\left(\phi^{m-1}\exp \left(-\frac12\phi^m\right)\right)
\end{equation}
This can integrated using the identity
\begin{equation}
\frac{d}{dx} (2m^{-1}x^{2-2m}e^{x^m/2})=x^{1-m}e^{x^m/2}\left(1+\frac{4-4m}{m}
x^{-m}\right)
\end{equation}
to give
\begin{equation}
t=\frac{2}{m^2}\sqrt{\frac{24\pi}{V_0}}\phi^{2-2m}\exp 
\left(\phi^m/2\right)(1+o(1)).
\end{equation}
This implies that
\begin{equation}
\phi^m=2\log t-(4-4m)\log\phi+2\log\left(\frac{m^2}{2}\sqrt{\frac{V_0}{24\pi}}
\right)+o(1)
\end{equation}
Taking logarithms gives
\begin{equation}
\log\phi=\frac1m\left(\log\log t+\log 2\right)+o(1)
\end{equation}
Substituting this back into the expression for $\phi^m$ gives
\begin{equation}
\phi^m=2\log t-\frac{4(1-m)}{m}(\log\log t+\log 2)+2\log\left(\frac{m^2}{2}
\sqrt{\frac{V_0}{24\pi}}\right)+o(1)
\end{equation}
Now the asymptotic form of the potential can be computed.
\begin{equation}
V(\phi(t))=V_0\left[\log\left(\frac{m^2}{2}\sqrt{\frac{V_0}{24\pi}}
\right)\right]^2 t^{-2}(2\log t)^{4(1-m)/m}(1+o(1))
\end{equation}
From this the asymptotic form of $H$ can immediately be computed
$H$.
\begin{equation}
H(t)=(8\pi V_0/3)^{1/2}\log\left(\frac{m^2}{2}\sqrt{\frac{V_0}{24\pi}}
\right) t^{-1}(2\log t)^{2(1-m)/m}(1+o(1))
\end{equation}

To end the section an example will be presented of a potential which does
not satisfy the assumptions of Theorem 4. This is given by:
\begin{equation}
V(\phi)=\phi^{-1}+\phi^{-7/2}\sin(\phi^2)
\end{equation}
In this case $V'(\phi)<0$ for $\phi$ large, $V'/V\to 0$ as $t\to\infty$,
$V''/V\to 0$ as $t\to\infty$ but $V''/V'$ is unbounded as $t\to\infty$.

\section{Specific matter models - perfect fluids and collisionless matter}

All the theorems of the previous sections have been obtained under the
assumption that a solution exists globally in the future. If a suitable
well-behaved model is chosen for the matter other than the scalar field 
then global existence for arbitrary initial data can be proved. In the
case of a perfect fluid with a linear equation of state this has been
done in \cite{rendall04a}. In the case of collisionless matter described 
by the Vlasov equation it has been done in \cite{lee04b}.

Once global existence has been shown it is possible to obtain detailed
information about the asymptotics of the matter fields. Consider an
untilted perfect fluid with linear equation of state $p=(\gamma-1)\rho$.
The state of the fluid is described entirely by its energy density $\mu$,
which satisfies $d\mu/dt=-3\gamma H\mu$. Let $l=(\det g)^{1/6}$. In the 
isotropic case $l$ agrees with the scale factor up to a multiplicative
constant. Now $dl/dt=Hl$. Hence $d/dt (l^{3\gamma}\mu)=0$ and the density
behaves like $l^{-3\gamma}$. From an asymptotic expression for $H$ we can 
obtain a corresponding expression for $l$ and hence an asymptotic expression 
for $\mu$. For example, for the solutions whose asymptotic behaviour was
analysed in the last section the leading order term in $l$ is proportional 
to $\exp ((\log t)^{(2-m)/m})$.

Consider now the case of collisionless matter with a potential satisfying the 
conditions of part a) of Theorem 4. 
\begin{equation}
\frac{d}{dt}\left(\frac{g_{ij}}{l^2}\right)=-2\left(\frac{g_{is}}{l^2}\right)
\sigma^s{}_j
\end{equation}
Now $\sigma_{ij}\sigma^{ij}$ decays faster than any power of $t$ and is,
in particular, integrable in $t$. It is possible to argue just as in
\cite{lee04a} to conclude first that $g_{ij}/l^2$ and its inverse are 
bounded and then that $g_{ij}/l^2$ converges as $t\to\infty$ to some
quantity $g^0_{ij}$. It follows that
\begin{equation}
g_{ij}=l^2(g_{ij}+h_{ij})
\end{equation}
where $h_{ij}=O(t^{-n})$ for any $n$. It is then possible to proceed as 
in \cite{lee04a} to obtain asymptotics for geodesics, i.e. for the 
characteristics of the Vlasov equation. The results are strictly analogous 
to those in \cite{lee04a} except that $e^{-Ht}$ is replaced by $l^{-n}$ in 
the error terms in \cite{lee04a}, where $n$ is arbitrary. For a specific 
potential where the form of $l$ can be determined these error 
estimates can be made sharper. If $V_i$ are the spatial components of
a tangent vector to a timelike geodesic then they converge faster than
any power of $t$ to a constant as $t\to\infty$. The energy density
decays like $l^{-3}$ as in the case of dust and all other components 
of the energy-momentum tensor in an orthonormal frame decay faster.

\section{Conclusion} 
Combining the results of this paper with previous work gives a rather
detailed picture of the dynamics of spatially homogeneous spacetimes with 
accelerated expansion driven by a minimally coupled nonlinear scalar field 
with potential. Potentials with a positive lower bound are covered in 
\cite{rendall04a}, potentials which are positive and tend to zero at 
infinity slower than exponentially are treated in this paper. Exponential 
decay with sufficiently small exponent has been handled in \cite{kitada92}, 
\cite{kitada93} and \cite{lee04b}. In the borderline case where the potential 
asymptotically approaches an exponential one but is not exactly exponential 
it would be desirable to get more information. In this paper accelerated 
expansion was demonstrated for this case for a scalar field alone in a
spatially flat isotropic model but isotropization in more general homogeneous
models, which is presumably true, was not. If the potential is zero somewhere 
then the expected behaviour is very different. In that case accelerated 
expansion is a transient phenomenon and the late time behaviour must be 
expected to depend in a complicated way on the Bianchi type. Cf. the 
discussion in the last section of \cite{rendall04a}.

The questions of global existence and the asymptotics of matter fields have 
been answered in some special cases. A treatment of more features of their 
asymptotic behaviour and of wider classes of matter fields would be 
desirable. Another direction in which the results should be generalized 
is to other mechanisms for producing accelerated expansion such as 
$k$-essence \cite{armendariz}. See also the discussion in \cite{rendall04b}.

For potentials which do lead to late-time accelerated expansion an
obvious next step is to look at inhomogeneous models. Expansions for
spacetimes without symmetry were written down for the case of a 
cosmological constant by Starobinsky \cite{starobinsky} and in the case of 
exponential potentials in \cite{mueller}. In \cite{rendall04c} the results 
of \cite{starobinsky} were given a rigorous mathematical interpretation
in terms of formal series and the existence of a large class of solutions
(depending on the maximum number of free functions) was demonstrated in
the vacuum case. Some of these results have been generalized to the case
of a potential with a positive minimum by Bieli \cite{bieli}. In the
vacuum case it was shown with the help of results of \cite{friedrich}
that all solutions evolving from small perturbations of de Sitter initial
data have late-time asymptotic expansions of the form given in 
\cite{starobinsky}. Presumably a corresponding result could be obtained
in any even spacetime dimension using the results of \cite{anderson}

For certain classes of spacetimes with high symmetry (spherical, plane and
hyperbolic symmetry) it has been shown in \cite{tchapnda1} and 
\cite{tchapnda2} that for many solutions of the Einstein-Vlasov 
system with positive cosmological constant there is late-time accelerated 
expansion and the geometry has leading-order asymptotics as in 
\cite{starobinsky}. It would be interesting to see to what extent these
results can be generalized to nonlinear scalar fields with potentials
which give accelerated expansion in the homogeneous case. The results
of \cite{tegankong} on the Einstein-Vlasov system coupled to a linear
scalar field are a first step in this direction. Note that the use 
of collisionless matter in these results is not an accident. If dust
or perfect fluid with pressure were used then it is to be expected that 
there would be no global existence of classical solutions. Cf. the  
results of \cite{rendalldust} and \cite{stahl}. The effect
of nonlinear scalar fields on cosmological expansion in the inhomogeneous
case has been studied heuristically and numerically in \cite{goldwirth}. 
One issue of particular interest which has no analogue in the homogeneous 
case is the formation of domain walls for potentials with more than one 
minimum.

The way in which information about inhomogeneities in the universe is
compared with theoretical models is via linear perturbation theory.
These calculations are at present out of reach of rigorous mathematical 
theorems. There is not a single result proving that an inhomogeneous 
solution of the full nonlinear equations is well approximated by the
expressions coming from linear perturbation theory.

Cosmological solutions of the Einstein equations with accelerated expansion
are a rich source of problems for mathematical relativity. It is to be hoped
that the mathematical theory will soon have reached the point where it
can contribute insights for the astrophysical applications of these 
solutions.

\end{document}